\documentstyle[12pt,epsf,aaspp4]{article}

\input{epsf}

\begin{document}


\def\figinsert#1#2{\epsfbox{#1} \message{#2} }          
\def\la{\lower.5ex\hbox{$\; \buildrel < \over \sim \;$}}
\def\ga{\lower.5ex\hbox{$\; \buildrel > \over \sim \;$}}
\def\r{\hangindent=1pc \noindent}
\def \ang{{\rm \AA}}
\def\sc{\scriptscriptstyle}
\def\h{H~I}
\def\he{He~I}
\def\hei{He \,II}
\def\12{{1\over 2}}
\def\msun{M_{\odot}}
\def\lsun{L_{\odot}}
\def\div{\nabla\cdot}
\def\grad{\nabla}
\def\rot{\nabla\times}
\def\eg{{\it e.g.,~}}
\def\ie{{\it i.e.,~}}
\def\etal{{\it et~al.,~}}
\def\de{\partial}
\def\ltsima{$\; \buildrel < \over \sim \;$}
\def\simlt{\lower.5ex\hbox{\ltsima}}
\def\gtsima{$\; \buildrel > \over \sim \;$}
\def\simgt{\lower.5ex\hbox{\gtsima}}
\def\noi{\noindent}
\def\bs{\bigskip}
\def\ms{\medskip}
\def\ss{\smallskip}
\def\ob{\obeylines}
\def\l{\line}
\def\hrf{\hrulefill}
\def\hf{\hfil}
\def\q{\quad}
\def\qq{\qquad}
\renewcommand{\deg}{$^{\circ}$}
\newcommand{\um}{$\mu$m}
\newcommand{\uk}{$\mu$K}
\newcommand{\qrms}{$Q_{rms-PS}$}
\newcommand{\n}{$n$}
\newcommand{\cdmr}{${\bf c}_{\rm DMR}$}
\newcommand{\xrms}{$\otimes_{RMS}$}
\newcommand{\gt}{$>$}
\newcommand{\lt}{$<$}
\newcommand{\ldl}{$< \delta <$}
\newcommand{\be}{\begin{equation}}
\newcommand{\ee}{\end{equation}}
\newcommand{\ba}{\begin{eqnarray}}
\newcommand{\ea}{\end{eqnarray}}


\title{Bubbles in galactic haloes}
 
\author{Yu. A. Shchekinov$^{1,2}$, R.-J. Dettmar$^2$, 
A. Schr\"oer$^2$, A. Steinacker$^{2,3}$}
\bs
{$^1$Department of Physics, Rostov State University,  
344090 Rostov on Don, Russia\\
\par
$^2$Astronomisches Institut, Ruhr-Universit\"at Bochum, 
D-44780 Bochum, Germany\\
\par
$^3$NASA Ames Research Center, Moffet Field, CA 94035, USA
}
\bs

     \date{Received ???, ???; accepted ???}


 
\begin{abstract}
We briefly discuss a possible interconnection of vertical HI structures 
observed in the Milky Way Galaxy with large scale blow-outs caused by 
the explosions of multiple clustered SNe. We argue that the observed OB 
associations 
can produce only about 60 such events, or approximately one chimney per 
3 kpc$^2$ within the solar circle. We also discuss the overall properties of 
HI shells in nearby face-on galaxies and the distribution of H$\alpha$ and 
dust in edge-on galaxies. We argue that the presence of dust in galactic haloes 
may indicate that radiation pressure is the most probable mechanism capable of 
transporting dust to large heights above the galactic plane. 
In order to make this possible, the galactic magnetic field  
must have a strong vertical component. We mention that SNe explosions can 
initiate the Parker instability which in turn creates large scale magnetic 
loops with a strong vertical component. Recent observations of nearby edge-on 
galaxies favour this suggestion. 

\end{abstract}



\section{Introduction}

Bubbles and shells in the interstellar gas can be discriminated by their sizes 
in three classes: bubbles with a characteristic size of 10 pc, superbubbles 
of about 100 pc in size, and supershells which extend upto 1 kpc or more.
This classification reflects the distinguishing
characteristics of the
dynamics of the expanding gas on scales of the local 
environment ($\sim 10$ pc, corresponding to a parent cloud where a star 
drives a wind or explodes as a SN), on characteristic scales of cold thin 
interstellar discs ($\sim 100$ pc), and on the much larger scales of gaseous 
haloes ($\sim 1$ kpc). In this contribution, we will 
concentrate on large scale structures, which extend far from 
the galactic planes -- the supershells. Our knowledge 
about these structures stems from observations of the gas distribution 
and its kinematics in the Milky Way and in nearby face-on and edge-on galaxies. 

\section{The Milky Way Galaxy}

In our Galaxy evidence for the existence of large scale shells extending far 
above the galactic plane comes from observations of neutral gas structures 
oriented predominantly in vertical direction and reaching heights of 200 pc 
to 1 kpc, the so-called worms (Heiles 1984). The most obvious and simple 
interpretation of worms is to regard them as vertical walls of chimneys -- 
large scale vertical outflows produced by clustered SNe explosions (Norman 
\& Ikeuchi 1989). (A less obvious and completely unexplored possibility can 
be connected with hydrodynamical instabilities producing vortices and 
vertical gas outflows similar to a tornado in the atmosphere). 
Following numerical simulations (see Mac Low \etal 1989, Silich \etal 1996), 
one can assume, that the total energy required for an expanding supershell to 
reach  a distance of $1-1.5$ 
kpc above the plane is of the order of $\sim 1-3\times 10^{53}$ erg. 
Then for a power law distribution of OB associations in the Galaxy 
(Kennicutt \etal 1989, Heiles 1990, Williams \& McKee 1997) 
\be
N_a(L)=5.5\left({475\over L_{49}}-1\right),
\ee
where $N_a(L)$ is the number of associations with an ionizing photon luminosity 
larger than $L$, $L_{49}=L/(10^{49}~{\rm s^{-1}})$, one obtains the total 
number of chimneys $N_{\rm ch}\simeq 60$, where $L_{49}=0.2$ for an O9 star 
is assumed (Shchekinov 1996). Thus for a total number of chimney walls 
seen in the projection we get $\simeq 120$ which is consistent with the 
observed number of worms $N_{\rm w}=118$ (Koo \etal 1992). Approximately half 
of the observed worms are associated with HII regions which very likely 
contain clustered SNe. In the infrared all the worms show a sufficiently 
large  
ratio $I(60~\mu {\rm m})/I(100~\mu {\rm m})\simeq 0.28$ (Koo \etal 1992) 
which can be connected with an excess of small grains (indicating probably 
that the material of worms has been processed by destructive shock waves). 
Apparantly, the closest chimney was detected in Cas OB6 association in
the Perseus arm (Normandeau \etal 1996). Assuming the distance to Cas OB6
$\sim 2.2$ kpc, one obtains the number of chimneys in the disc $N_{\rm
ch}\sim (15~{\rm kpc/2.2~kpc})^2\simeq 46$ which is consistent with the
above estimates. 
These arguments favour the interpretation of worms as chimneys or supershells 
driven from the galactic plane by SNe explosions. 

\section{Face-on galaxies}

Large scale HI shells in nearby face-on galaxies have been
known for almost 20 years 
since the first detection in M31 (Brinks, 1981). Since then HI shells 
have been studied in detail in several tens of 
face-on galaxies -- besides M31 the most interesting of these are: 
M33 (Deul \& den Hartog 1990), Holmberg II (Puche \etal 1992), M101 and 
NGC 6496 (Kamphius 1993), SMC (Staveley-Smith \etal 1997), 
IC 10 (Wilcots \& Miller 
1998), LMC (Kim \etal 1998, 1999), IC 2574 (Walter \& Brinks 1999). Typically 
50 to 100 shells with sizes ranging in the interval 100 pc -- 1 kpc 
have been observed, 
and this allows simple statistical conclusions. Three quantities are observed: 
the radius $R$ (more precisely the angular diameter) of a shell in the 
galactic plane, the expansion velocity $v$ along the line of sight, and the 
column density of HI, $\sigma(HI)$, in a
close vicinity of the shell; the volume density $n(HI)$ can then be 
inferred from $\sigma(HI)$ and the estimated scale height (for details see: 
Puche \etal 1992, Walter \& Brinks 1999). When the observed points are plotted 
in the radius--velocity plane, no obvious correlation is seen. They are distributed
randomly (Fig. 1, here data for HoII, LMC and IC 2574 are plotted). 
However, the quantity $nv^2R^3$ (which is, as is readily seen,  
the explosion energy $E$ calculated under the assumption that the expansion 
always remains adiabatic) shows a clear correlation (see Fig. 2 where 
on the vertical axis $E^{1/3}=[nv^2R^3]^{1/3}$ is plotted). 
This correlation justifies 
the assumption that regardless of the origin of the correlation,
all the shells are produced by SN energy input. 
(Note, that this correlation means merely 
$nv^2\simeq$ constant, which corresponds to pressure modified stages of 
expansion when the ram pressure is close to the external pressure. From this 
point of view the scatter is partially due to the fact that the expansion velocities 
are not exactly equal to the sound speed). However, one feature captures the 
attention: the velocities corresponding to the holes in HoII are mostly 
subsonic. It is easily seen from Fig. 1 that approximately 75\% of the points 
lie below the sound speed (which for HoII is $\simeq$ 8 km s$^{-1}$). Half 
of the subsonic shells have ages of about 100 Myr or more. Thus, three questions 
have to be answered: how can the subsonic shells (whose dynamics therefore is 
strongly modified by the external pressure) possibly follow approximately 
the same trend 
as the supersonic shells in IC 2574 do, how do they keep their integrity 
during such a long time and why are they not destroyed by external turbulent 
motions? 
Note that turbulence destroys subsonic shells with a characteristic 
time $t_{\rm d}\sim R/c_s$ which is less than the estimated age $\sim 
R/v$. It is worth stressing that at late stages radiation pressure acting 
on the expanding shells can be dynamically important. For a typical value 
of the interstellar radiation energy flux $\Phi \sim 10^{-2}$ erg cm$^{-2}$ 
s$^{-1}$, the radiation force per unit volume is 
$f_R\sim 3\times 10^{-34}\xi n$ cgs, where $\xi$ is the dust-to-gas ratio 
normalized to its value in the local ISM and $n$ is the mean density in the shell. 
At the same time, the gradient of thermal pressure in the remnant is 
$\nabla p\sim E/2\pi R^4\sim 5\times 10^{-34}$ cgs, $E=3\times 10^{53}$ erg is the 
explosion energy, $R\sim 1$ kpc, its radius (Puche \etal 1992). It is seen that 
for $n\sim 1-3$ cm$^{-3}$, $f_R$ is comparable to $\nabla p$, and it can 
be even larger if the dust-to-gas ratio $\xi$ is enhanced in haloes as suggested 
by Dettmar \etal (2001).

The mass and energy distribution functions of the shells are usually 
peaked, with a maximal number of shells at $M\sim 10^5-10^6~\msun$, and 
at $E\sim 1-3\times 10^{51}$ erg for different galaxies (see discussion in 
Walter \& Brinks 1999). The HI spatial resolution (around $\sim 100$ pc) 
corresponds to considerably smaller masses of $\sim 3-10\times 10^3~\msun$, 
and thus the decline in the mass (and correspondingly, in the energy) 
distribution at low masses cannot be attributed to the spatial resolution. 
Since these distributions peak at energies corresponding to a few SNe,
the decline can be connected with the cutoff in the distribution of OB 
associations at the low luminosity end. The decline in distributions 
at higher energies and masses can reflect a universal power-law luminosity 
function of OB associations. 

Walter \etal (1998) have detected soft X-ray emission 
from a region coincident with an HI supershell in IC 2574. Its spectrum agrees 
with a Raymond-Smith spectrum at $T=10^{6.8}$ K indicating that combined 
effects of stellar winds and sequential SNe explosions are the energy source 
for this region (although a contribution 
from binaries cannot be excluded). However, the expansion velocity of the 
HI shell, $\sim 25$ km s$^{-1}$, suggests rather late stages of the bubble 
when its temperature is at least ten times smaller, $\sim 3-5\times 10^5$ K
(see, Slavin \& Cox 1992). A possible explanation for this disagreement 
can be based on a SN explosion which has occured recently
($t\ll 10$ Myr) so that the blast wave has not yet reached the expanding HI 
shell.    

\section{Edge-on galaxies}

Edge-on galaxies provide us with direct information about the vertical 
distribution of 
gas, cosmic rays, and magnetic fields in galactic haloes (Dettmar 1992). 
Recent high-resolution observations with unsharp-masked technique have 
shown the presence of highly organized dust features far from the galactic planes 
of edge-on galaxies. E.g., Sofue \etal (1994) have found multiple 
vertical ``dust streamers'' in the nearly edge-on galaxy NGC253 
(inclination 78\deg) extending coherently up to 1 - 2 kpc. 
Howk \& Savage (1997, 2000)  
have obtained deep optical images in BVI and H$\alpha$ 
of the edge-on galaxy NGC891. They found that dust and ionized gas extend up
to distances $|z|\sim 2$ kpc from the midplane. Dust extinction and H$\alpha$ 
emission are found to have a filamentary and clumpy structure, however without 
a direct physical relationship between the two components. In most cases at 
lower heights ($|z|\simlt 1$ kpc) the structure seen in H$\alpha$ is mostly 
due to absorption by dust-bearing clouds (Howk \& Savage 2000). This probably 
suggests that dust features and H$\alpha$ gas have a distinct origin. The 
presence of dust may indicate that commonly discussed strong blowouts from SNe 
cannot be considered as a dominant transport mechanism of material 
in vertical direction. For a blowout to occur, the SN shock wave must always be 
supersonic, which means that when the shock enters the predominantly hot phase 
($T\sim 10^6$ K) at $|z|\sim 0.4-0.5$ kpc, its velocity must be $v_{\rm sh}>
100$ km s$^{-1}$. However, when being processed by such shocks dust grains are 
easily destroyed (Drain 1995). More detailed calculations show that up to 
30-50 \%  are destroyed by such shock waves (Dettmar \etal 2001). 
Therefore, ``soft'' transport mechanisms are needed to provide for an elevation 
of dust far above the galactic planes. 

Among such mechanisms (see detailed discussion in Howk \& Savage 1997), the 
radiation pressure of stellar light acting on dust particles is considered as 
effective (Barsella \etal 1989, Franco \etal 1991, Ferrara \etal 1991, 
Ferrara 1998). However, because dust particles are charged (carrying a positive 
charge when immersed in a diffuse gas) they are strongly coupled to the
magnetic field -- the gyro-radius is about $r_G\sim 3\times 10^{11}v_{\rm d}$ 
for $B\sim 3~\mu$G, and grain radius $a\sim 0.1~\mu$m, $v_{\rm d}$ is the 
velocity of a dust particle in cgs; for subsonic particles $r_G$ is less than 
1 pc. In principle, they can spend intermittently some time in neutral states 
due to charge fluctuations as mentioned by Ferrara (1998), however the 
characteristic charging time is normally much shorter than 1 yr, and the 
contribution of such intermittent periods is negligibly small. 
Thus, in order to be efficient, this mechanism requires the
presence of a locally strong vertical magnetic field
component in the interstellar medium. A possible mechanism to 
produce such a configuration is the Parker instability
(Sokoloff \& Shukurov, 1990). Recent simulations (Kamaya et al.,
1996, Steinacker \& Shchekinov, 2001) show that the instability
can be triggered by SN explosions even if their total energy
release is moderate (E $\ll 10^{53}$ erg). However, Steinacker \& Shchekinov
have shown that the time scale on which significantly large
loops are being formed, depends on the magnitude of the
gravitational acceleration $g$. Only for $g\geq 4.5\times 10^{-9}$  
cm s$^{-2}$, do
the loops become sufficiently prominent within a characteristic
time comparable to the rotation period of the galactic disc. In these
cases, the loops can extend up to 2-3 kpc into the halo (Fig. 3).
The production of such prominent loops can be facilitated, if the
Parker instability and the SN remnant expansion
can operate simultaneously and interact. In order
for this interaction to take place, the energy provided by the explosion
must be larger than the minimal energy required for the Parker
instability to be initiated and lower than the minimal energy required
for a blow-out to occur. In this case the instability evacuates
gas from the expanding shell, therefore facilitating the expansion
of the interior hot bubble, or the expanding bubble carries material
away, allowing for the magnetic field to rise faster. Furthermore,
they have shown that even one SN explosion can generate multiple loops.
These secondary loops are a consequence of the fact that the
perturbation induced by the explosion propagates through the disc. 
As seen from the above estimates 
of the radiation pressure, a local increase of the radiation flux 
from a young OB association can also initiate the Parker instability. 

Once formed, a magnetic loop can serve as a conductor for radiatively driven 
dust particles. At heights $z\sim 200$ pc collisional coupling between the 
dust and gas weakens because of the exponential decrease of the gas density, and 
at higher $z$ dust moves practically friction-free so that the dust-to-gas ratio
gets enhanced. At heights $z\sim 1.5$ kpc the friction between the dust and 
gas is so weak that dust particles oscillate along the magnetic arch, and 
form a horn-like density distribution as shown in Fig. 4 (Dettmar \etal 2001).  
The peak density of dust in the clumps is around 20 times the midplane value, 
and as dust and gas are dynamically weakly coupled at these heights, actual 
dust-to-gas ratios can be considerably higher. 

\section{Summary.} 

The total number of chimneys in the Milky Way produced by clustered SN
explosions with the observed luminosity function for OB associations is in 
agreement with the total number of worms (Koo \etal 1992) and the chimneys 
detected in Cas OB6 association (Normandeau \etal 1996).

The presence of subsonic HI shells in HoII can be connected with the action of 
radiation pressure at late stages of the expanding SN remnants. 

The underlying cause for the observed presence of dust in the haloes of
edge-on galaxies may be explained as a combined effect of magnetic fields
and radiation pressure acting to transport matter vertically. In order for
this trasnport mechanism to occur, the interstellar magnetic field must
be reorganized by the Parker instability, hence gaining a considerable
vertical component. The Parker instability, in turn, can be initiated by a
relatively small amount of energy released from the SN explosions.

\begin{acknowledgements} 
This work was supported by the Deutsche For\-schungs\-ge\-mein\-schaft
(Bonn, Germany) through Sonderforschungsbereich 191 "Phy\-si\-ka\-li\-sche
Grundlagen der Nie\-der\-tem\-pe\-ra\-tur\-plas\-men". 
YS acknowledges partial support from Federal 
Programme ``Astronomy''. 
\end{acknowledgements} 

\bs


\begin{figure}
\epsfxsize=15.5truecm   
\epsfysize=15truecm     
\epsfbox{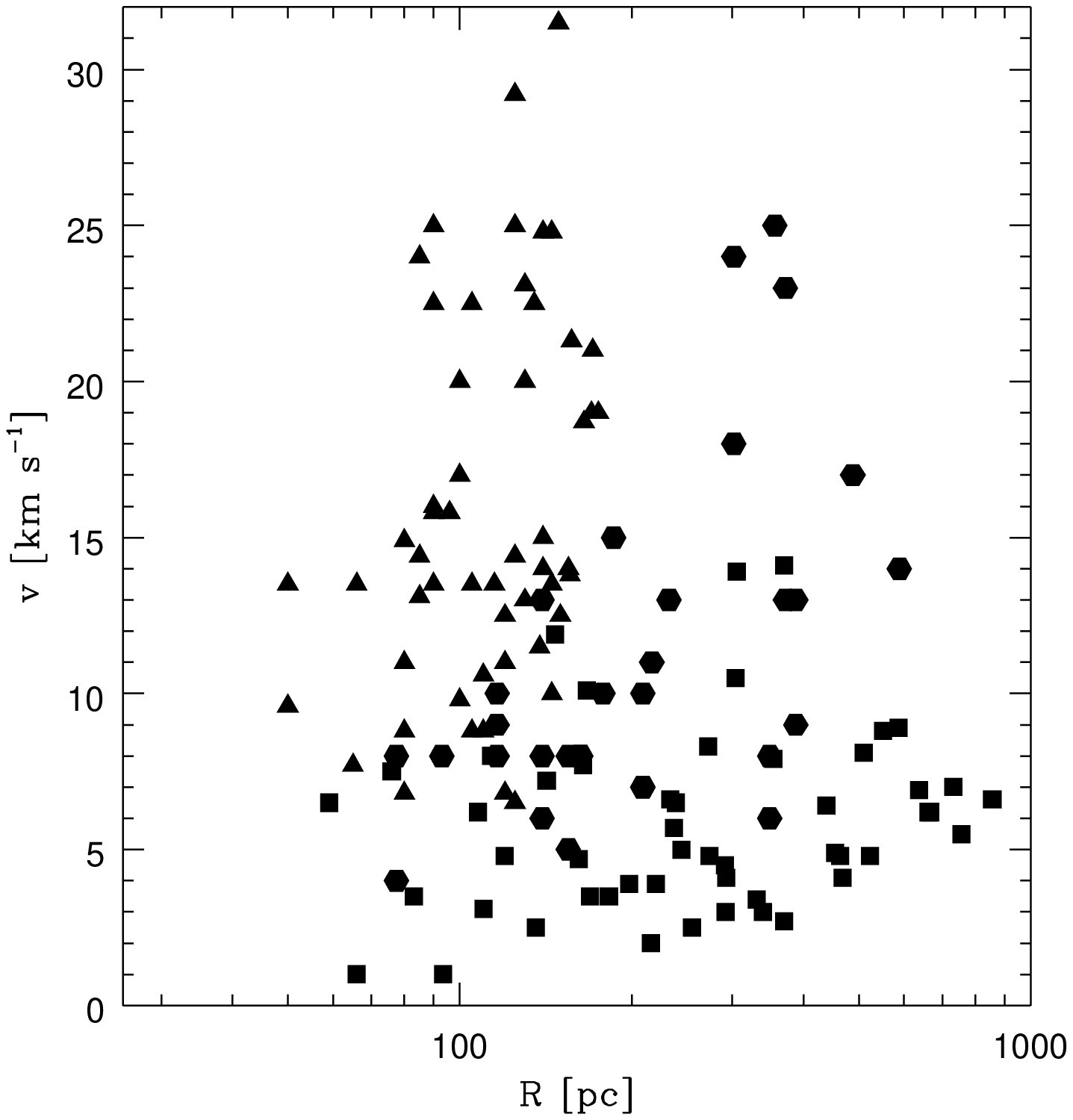}
\caption{{Expansion velocities of the HI shells versus their radius 
observed in three face-on galaxies: LMC (triangles), HoII (squares), 
IC 2574 (hexagons) from Kim \etal 1999, Puche \etal 1992, and 
Walter \& Brinks 1999, respectively. 
}}
\end{figure}

\begin{figure}
\epsfxsize=15.5truecm
\epsfysize=15truecm
\epsfbox{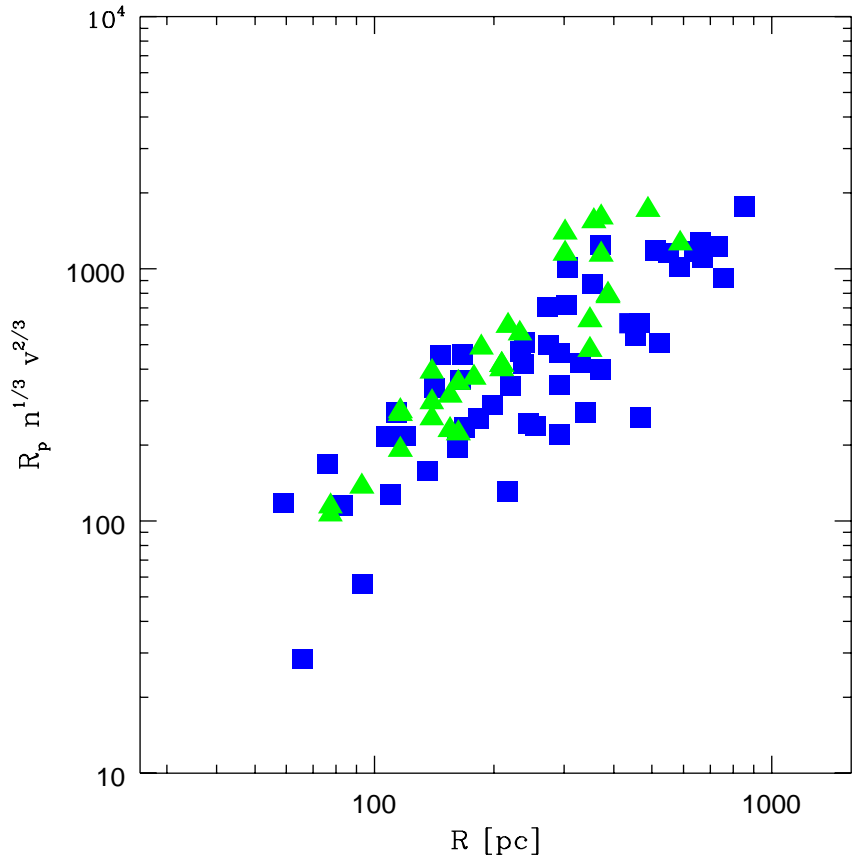}
\caption{{$E^{1/3}=Rn^{1/3}v^{2/3}$ (see text) versus radius for 
IC 2574 (triangles), and HoII (squares). 
}}
\end{figure}

\begin{figure}
\epsfxsize=15.5truecm
\epsfysize=15truecm
\epsfbox{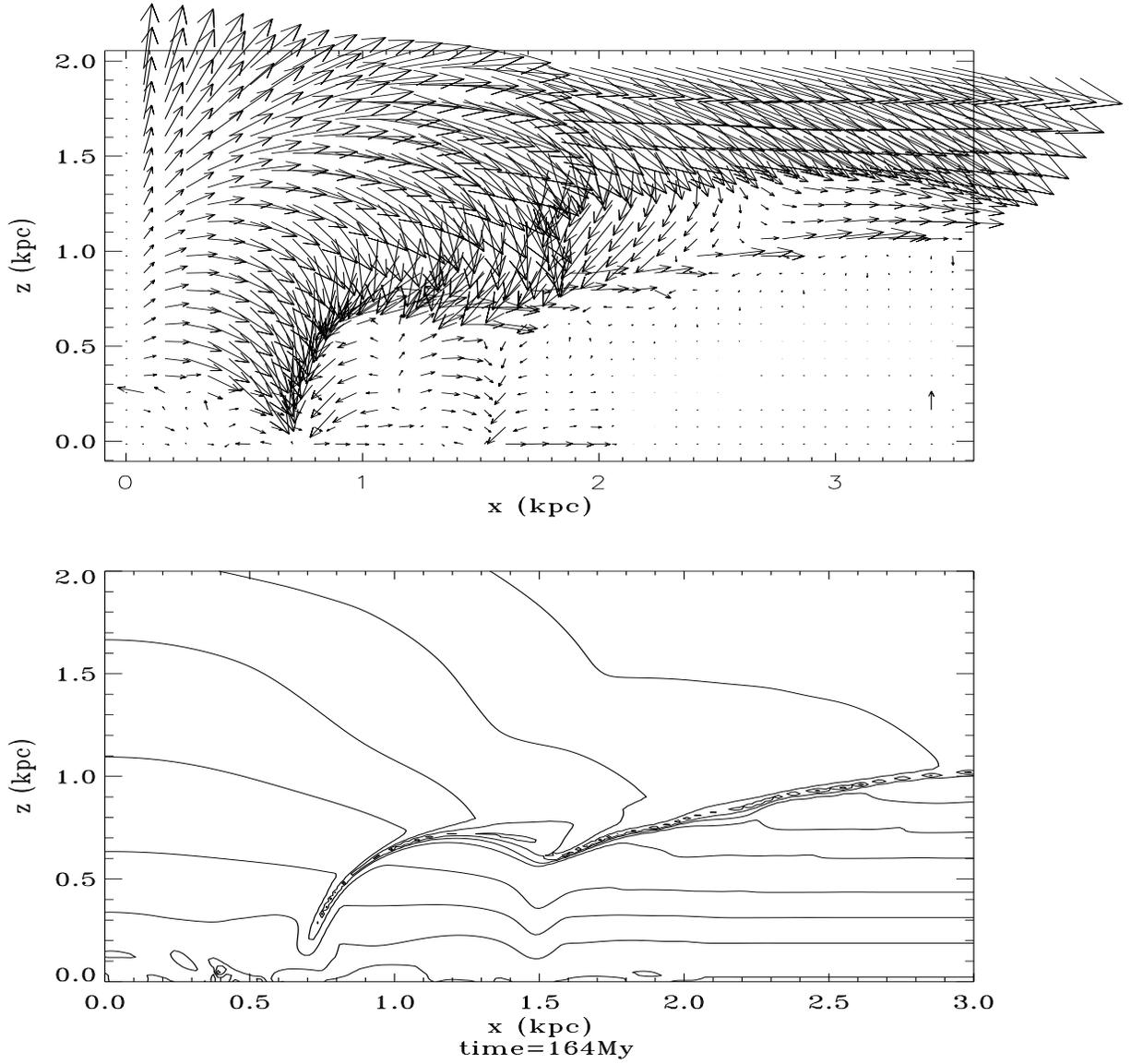}
\caption{{Magnetic lop formed by the Parker instability initiated by a SN 
explosion (in the origin) after 164 Myr (Steinacker \& Shchekinov 2000).
}}
\end{figure}

\begin{figure}
\epsfxsize=15.5truecm
\epsfysize=15truecm
\epsfbox{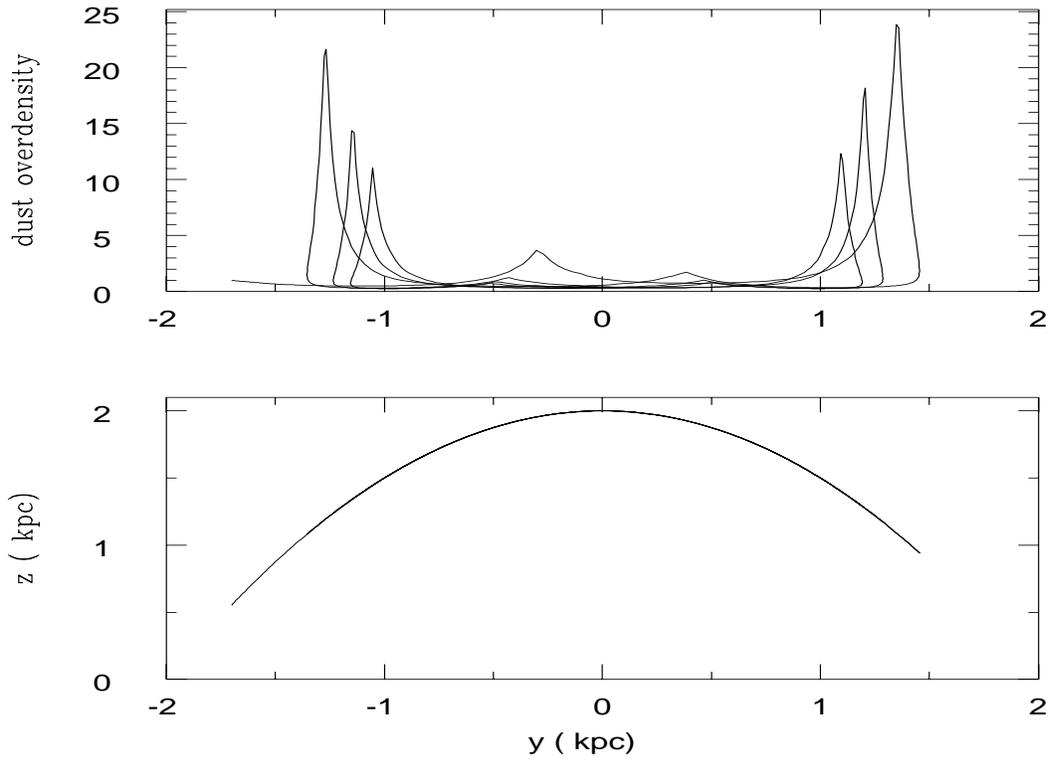}
\caption{{Dust density distribution (upper panel) formed by the radiation 
pressure driving the dust particles along a magnetic arc (lower panel). 
}}
\end{figure}

\end{document}